\documentclass[%
reprint,
bibnotes,
amsmath,amssymb,
aps,
prl,
one column
]{revtex4-2}

\usepackage{graphicx}
\usepackage{dcolumn}
\usepackage{bm}
\usepackage{mathrsfs}
\usepackage{comment}
\usepackage[usenames,dvipsnames]{color}
\usepackage{xcolor}
\usepackage{slashed}
\usepackage{cancel}
\usepackage{simplewick}
\usepackage{appendix}

\usepackage{tikz}
\usetikzlibrary{arrows,shapes}
\usetikzlibrary{trees}
\usetikzlibrary{matrix,arrows}
\usetikzlibrary{positioning}				
\usetikzlibrary{calc,through}				
\usetikzlibrary{decorations.pathreplacing}  
\usepackage{pgffor}							

\usetikzlibrary{decorations.pathmorphing}	
\usetikzlibrary{decorations.markings}
\tikzset{
	doublearrow/.style={draw=black,double =black,double distance=2.0pt,
		postaction={decorate},decoration={markings,mark=at position 1.0 with {\arrow[draw=black,line width=1.5pt]{>}}}},
	heavyquark/.style={draw=black,double =white,double distance=2.0pt,
		postaction={decorate},
		decoration={markings,mark=at position .55 with {\arrow[draw=black,line width=1.5pt]{>}}}},
	heavyquarkbar/.style={draw=black,double =white,double distance=2.0pt,postaction={decorate},
		decoration={markings,mark=at position .55 with {\arrow[draw=black,line width=1.5pt]{<}}}},
	solidline/.style={draw=black, postaction={decorate}},
	thickline/.style={draw=black, line width=4.0pt, postaction={decorate}},
	doubleline/.style={draw=black,double =white,double distance=2.0pt,postaction={decorate}},
	photon/.style={decorate, decoration={snake,amplitude=2.0pt, segment length=4pt}, draw},
	prophoton/.style={decorate, decoration={snake,amplitude=2.5pt}, draw},
	antiphoton/.style={decorate, decoration={snake,amplitude=-2.5pt}, draw},
	fermion/.style={draw=black, postaction={decorate},
		decoration={markings,mark=at position .5 with {\arrow[draw=black]{>}}}},
	antifermion/.style={draw=black, postaction={decorate},
		decoration={markings,mark=at position .5 with {\arrow[draw=black]{<}}}},
	fermionnoarrow/.style={draw=black},
	gluon/.style={decorate, draw=black,
		decoration={coil,amplitude=2.5pt, segment length=3.5pt}},
	ghost/.style={dashed,draw=black, postaction={decorate},
		decoration={markings,mark=at position .55 with {\arrow[draw=black]{>}}}},
	antighost/.style={dashed,draw=black, postaction={decorate},
		decoration={markings,mark=at position .55 with {\arrow[draw=black]{<}}}},
	scalarnoarrow/.style={dashed,draw=black},
	electron/.style={draw=black, postaction={decorate},
		decoration={markings,mark=at position .55 with {\arrow[draw=black]{>}}}},
	bigphoton/.style={decorate, decoration={snake,amplitude=4pt}, draw},
	arrow/.style={draw=black, postaction={decorate},
		decoration={markings,mark=at position 1. with {\arrow[draw=black]{>}}}},
}
\tikzstyle{block} = [draw, rectangle,
minimum height=3em, minimum width=6em]
\usepackage{hyperref}

\begin{document}
\title{Quantum color screening in external magnetic field}
\author{Guojun Huang}
\author{Jiaxing Zhao}
\author{Pengfei Zhuang}
\address{Physics Department, Tsinghua University, Beijing 100084, China}

\begin{abstract}
We calculate color screening mass in a thermalized and magnetized QCD matter in the frame of loop resummation theory without restriction to the magnetic field strength. Our full calculation covers the often used approximations for weak magnetic field at high temperature and strong magnetic field at low temperature. We find that while the magnetic field created in heavy ion collisions at RHIC and LHC energies is probably the strongest one in nature, its effect on the QCD matter is still weaker in comparison with the high temperature of the fireball, and therefore can safely be treated as a perturbation.   
\end{abstract}
\maketitle

\section{Introduction}
The color charge of a quark in a quark-gluon plasma (QGP) will be screened by the surrounding quarks, antiquarks and gluons. This phenomenon in strong interaction is in analogy to the well-known Debye screening of an electric charge in electrodynamic interaction. The screening strength is normally characterized by the Debye mass $m_D$ which is inversely proportional to the screening length $r_D$. When the distance between two colored quarks is larger than the screening length, the averaged color interaction between them disappears. For a quarkonium like $J/\psi$ in a QGP, the color screening effect reduces the potential between the pair of heavy quarks and leads to a quarkonium suppression in high energy nuclear collisions~\cite{Matsui:1986dk}. In the limit of high temperature QGP, the Hard-Thermal-Loop resummed perturbation theory~\cite{QFT} works well and gives an analytic Debye screening mass~\cite{Laine:2006ns,Schneider:2003uz}, $m_D^2(T)=(N_c/3+N_f/6)g^2T^2$, where the first term comes from gluons and ghosts and the second term is the contribution from massless quarks, $N_c$ and $N_f$ are the numbers of color and flavor degrees of freedom, and $T$ and $g$ are the plasma temperature and coupling constant in quantum chromodynamics (QCD). The calculation has been extended to any temperature, chemical potential and anisotropic medium~\cite{Huang:2021ysc,Dong:2021gnb,Thakur:2020ifi}.

Since the screening phenomenon happens in both strong and electromagnetic interactions, a natural question is the color screening in an external electromagnetic field. If the field strength is strong enough, its effect on the color screening may not be neglected. In fact, a hot QGP system under strong electromagnetic field is expected to be created in the early stage of high energy nuclear collisions at the Relativistic Heavy Ion Collider (RHIC) and the Large Hadron Collider (LHC), where the magnetic field may respectively reach $eB\sim 5m_\pi^2$ and $70 m_\pi^2$~\cite{Skokov:2009qp,Voronyuk:2011jd,WTDeng:2012prc,Tuchin:2013ie,Gursoy:2014aka,Yan:2021zjc,Chen:2021nxs,Wang:2021oqq} with $m_\pi$ being the pion mass in vacuum. This is probably the strongest magnetic field in nature and has led to a number of interesting discussions in the study of high energy nuclear collisions, for instance the effect of magnetic catalysis or inverse magnetic catalysis on QCD phase structure~\cite{Shovkovy:2012zn,Bruckmann:2013oba}, spin induced quantum fluctuations like chiral magnetic effect~\cite{Kharzeev:2008npa,Fukushima:2008xe}, splitting of $D$ and $\bar D$ directed flow~\cite{Adam:2019wnk,Acharya:2019ijj,Das:2016cwd}, and changes in quarkonium properties and distributions ~\cite{Marasinghe:2011bt,Alford:2013jva,Machado:2013rta,Cho:2014exa,Guo:2015nsa,Bonati:2015dka,Yoshida:2016xgm,Zhao:2020jqu,Mishra:2020kts,Chen:2020xsr,Iwasaki:2021nrz,ALICE:2015mzu,Zha:2018ytv,Klein:2018fmp}. On the electromagnetic effect on color screening, most of the studies are in the two limits of weak and strong magnetic field in comparison with the medium temperature. For the former one takes Taylor expansion of the field~\cite{Hasan:2020iwa,Karmakar:2018aig}, and for the latter the lowest-Landau-level approximation is used~\cite{Bonati:2017uvz,Singh:2017nfa,Hasan:2018kvx,Hasan:2017fmf}. In both cases the Debye screening mass $m_D$ increases with the magnetic field strength. The other question we ask ourselves is the quark energy quantization. As is well-known in non-relativistic quantum mechanics, the transverse energy of a free fermion in an external magnetic field $B$ is quantized as $\epsilon_n=(2n+1)|qB|/(2m)$~\cite{Landau:1930} with $n$ being a positive integer, $m$ the fermion mass and $q$ the fermion electric charge. Is there still this quantization in the quark loop calculation? 

Considering the fact that the magnetic field created in high energy nuclear collisions may not satisfy the conditions to be weak or strong with respect to the fireball temperature, it is necessary to go beyond the two limits and study the magnetic field effect on the color screening without restriction to the field strength. In this paper, we generally calculate the color screening mass in the frame of resummed QCD perturbation theory at finite temperature and magnetic field. We will firstly introduce the quark propagator in thermal and magnetized QGP, and then derive the gluon self-energy and in turn the color screening mass. We will focus on the process of how the quark transverse energy is quantized in the calculation of gluon polarization. We will finally come back to the well-known results in the two limits of weak and strong magnetic field. Since gluons do not interact directly with the external magnetic field, the field contribution to the gluon self-energy and screening mass arises only from the quark loop.

\section{Quark propagator}
From the minimum coupling principle, the quark propagator $G(x,x')$ in an external (classical) magnetic field along the $z-$axis ${\bm B}=B{\bm e}_z$ derivable from the potential $A_\mu$ is controlled by the equation of motion,
\begin{equation}
\left(i\gamma\cdot\partial+q\gamma\cdot A-m\right)G(x,x') =\delta(x-x').
\end{equation}

Introducing the kinematical momentum operator $\hat\Pi_\mu=\hat p_\mu+qA_\mu$, as distinguished from the canonical momentum operator $\hat p_\mu$, and solving the Dirac equation, the operator $\hat {\cal H}=-(\gamma\cdot\hat\Pi)^2 = -\hat\Pi^2-(q/2)\sigma^{\mu\nu}F_{\mu\nu}$ with $F_{\mu\nu}$ being the electromagnetic field tensor satisfies the eigen equation,
\begin{equation}
\hat {\cal H} \left|p\right> = \left(-p_0^2+m^2+p_z^2+\epsilon_{lm_lm_s}^2\right)\left|p\right>
\end{equation}
with the Landau energy levels~\cite{Landau:1930} $\epsilon_{lm_lm_s}^2=2l|qB|+\left[1-|m_l|-\text{sgn}(q)\left(m_l+2m_s\right)\right]|qB|$ characterized by the quantum numbers $l=0,1,\cdots,\infty, m_l=-l,\cdots,0,\cdots,l$ and $m_s=-1/2,1/2$ and the four-momentum eigen state $\left|p\right>=\left|p_0,p_z,l,m_l,m_s\right>$. It is clear that $\hat{\cal H}$ is the difference between $p_0^2$ and the on-shell energy square. For on-shell quarks the difference disappears, but in general state with arbitrary $p_0$ the Landau energy levels $\epsilon_{lm_lm_s}^2\geq 0$ and the constraint $-p_0^2\geq 0$ (in the imaginary time formalism of finite temperature field theory) lead to a positive definite $\hat{\cal H}$.

The quark propagator in the corresponding Euclidean space can be represented in terms of $\hat {\cal H}$~\cite{Chyi},
\begin{eqnarray}\label{Green_function_1}
G(x,x') &=& \left<x\right|{1\over \gamma\cdot\hat\Pi-m}\left|x'\right>\\
&=& -\int_0^\infty ds\left<x\right|(\gamma\cdot\hat\Pi+m)e^{-(m^2+\hat {\cal H})s}\left|x'\right>.\nonumber
\end{eqnarray}
Taking a transformation from the $s$-independent momentum $\hat \Pi$ to the $s$-dependent momentum $\hat \Pi(s)=\hat U(-s)\hat \Pi\hat U(s)$ through $\hat U(s) = e^{-\hat {\cal H} s}$, the quark propagator can be written as
\begin{equation}
\label{g}
G(x,x') = -\int_0^\infty ds\ e^{-m^2s}\left<x\right|\hat U(s)(\gamma\cdot\hat\Pi(s)+m)\left|x'\right>.
\end{equation}

Similar to the transformation from the Schr\"odinger picture to the Heisenberg picture in quantum mechanics, the $s$-dependence of the momentum and coordinate operators are controlled by the Heisenberg-like equations,
\begin{eqnarray}
	\partial_s\hat\Pi_\mu(s)&=& \left[\hat {\cal H}, \hat\Pi_\mu(s)\right]=2iqF_{\mu\nu}\Pi^\nu(s),\nonumber\\
	\partial_s\hat x_\mu(s) &=& \left[\hat {\cal H}, \hat x_\mu(s)\right]=-2i\hat \Pi_\mu(s)
\end{eqnarray}
which lead to the solutions
\begin{eqnarray}
\label{pi(s)}
	\hat \Pi(s) &=& e^{2i{\cal F}s}\hat\Pi(0),\nonumber\\
	\hat x(s) &=& \hat x(0)+{\cal N}(s)\hat \Pi(0)
\end{eqnarray}
with the two matrices ${\cal F}$ and ${\cal N}(s)$ defined as
\begin{eqnarray}
&& {\cal F} = q{F^\mu}_\nu = qB\left[\begin{matrix}
	0&0&0&0\\
	0&0&-1&0\\
	0&1&0&0\\
	0&0&0&0
\end{matrix}\right],\nonumber\\
&& {\cal N}(s) = -{1\over qB}\left[\begin{matrix}
	2isqB & 0 & 0 & 0 \\
	0 & i\sinh(2qBs) & -2\sinh^2(qBs) & 0 \\
	0 & 2\sinh^2(qBs)& i\sinh(2qBs) & 0 \\
	0 & 0 & 0 & 2isqB
\end{matrix}\right].
\end{eqnarray}

Combining the two solutions in (\ref{pi(s)}) together, the momentum operator is represented by the coordinate operator,
\begin{equation}
\hat\Pi(s) = e^{2i{\cal F}s}{\cal N}^{-1}(s)\left[\hat x(s)-\hat x(0)\right],
\end{equation}
and the first matrix element in the quark propagator (\ref{g}) becomes
\begin{equation}
\label{first}
\left<x\right|\hat U(s)\hat\Pi(s)\left|x'\right> = e^{2i{\cal F}s}{\cal N}^{-1}(s)(x-x')\left<x\right|\hat U(s)\left|x'\right>.
\end{equation}
The second matrix element satisfies the evolution equation
\begin{equation}
\partial_s\left<x\right|\hat U(s)\left|x'\right> = -\left<x\right|\hat U(s)\hat H\left|x'\right>
\end{equation}
which results in the solution
\begin{equation}
\label{second}
\left<x\right|\hat U(s)\left|x'\right> = {1\over 16\pi^2 s^2}e^{(x-x')^\mu K_{\mu\nu}(s)(x-x')^\nu-\ln\left[{\sinh(qBs)\over qBs}\right]+{1\over 2}q\sigma_{\mu\nu}F^{\mu\nu}s}
\end{equation}
with 
\begin{equation}
K_{\mu\nu}(s) = {qB\over 4}{\rm diag}\left({1\over qBs},-{1\over \tanh(qBs)},-{1\over \tanh(qBs)},-{1\over qBs}\right).
\end{equation}
In the limit of $s\rightarrow 0$, the matrix element $\left<x\right|\hat U(s)\left|x'\right>$ goes back to the delta function, $\lim_{s\rightarrow 0}\left<x\right|\hat U(s)\left|x'\right>=\delta(x-x')$.

Substituting the two matrix elements (\ref{first}) and (\ref{second}) into the quark propagator (\ref{g}), taking the replacements of $s\rightarrow -is$, and then introducing a dimensionless variable $v=|qB|s$, we obtain, after a Fourier transformation from coordinate space to momentum space, the quark propagator in the external magnetic field, 
\begin{equation}
\label{quark}
G(p) = -\int_0^\infty{dv\over|qB|}\left\{\left[m+(\gamma\cdot p)_{||}\right]\left[1-i{\rm sgn}(q)\gamma_1\gamma_2\tanh v\right]-{(\gamma\cdot p)_\perp\over\cosh^2 v}\right\}e^{-{v\over |qB|}\left[m^2-p_{||}^2+{\tanh v\over v} p_\perp^2\right]}.
\end{equation}
Since the magnetic field breaks down the translation invariance, the quark momentum $p$ is now separated into a longitudinal and a transverse part $p_{||}$ and $p_\perp$, parallel and perpendicular to the magnetic field. Note that, except for a Schwinger phase the propagator (\ref{quark}) is the same as originally derived by Schwinger 70 years ago~\cite{Schwinger:1951nm,Alexandre:2000jc}. Since the two phase factors for the quark and anti-quark of a loop will cancel to each other in the calculation of color screening mass, we will neglect the phase in the following. 

\section{Gluon polarization}
With the known quark propagator we can now calculate the gluon polarization function, namely the quark loop function $\Pi_{\mu\nu}(k)$. After the usually used summation over quark loops on a chain, one can derive a non-perturbative gluon propagator~\cite{QFT}. Since we are interested in the color screening mass which is determined by the pole of the gluon propagator, we will focus on the polarization in the limit of zero momentum $\lim_{{\bm k}\to 0}\Pi_{\mu\nu}(k_0=0,{\bm k})$~\cite{QFT}. Considering the fact that the thermal and magnetized medium does not bring in any new divergence in the field calculation, we directly calculate $\Pi_{\mu\nu}(0,{\bf 0})$ in the following, and explicitly express its temperature and magnetic field dependence as $\Pi_{\mu\nu}(T,B)$.

Using the invariance of the quark loop under the substitution of the integrated quark momentum $p_\mu\rightarrow -p_\mu$, the polarization can be simplified as
\begin{eqnarray}
\Pi_{\mu\nu}(T,B) &=& {g^2T\over 2|qB|^2}\sum_{n{\bm p}v_1v_2}{\rm Tr}\bigg\{{(\gamma\cdot p)_\perp\gamma_\mu(\gamma\cdot p)_\perp\gamma_\nu\over\cosh^2v_1\cosh^2v_2}+\left[1-i{\rm sgn}(q)\gamma_1\gamma_2\tanh v_1\right]
\left(m^2\gamma_\mu-\omega_n^2\gamma_0\gamma_\mu\gamma_0+p_z^2\gamma_3\gamma_\mu\gamma_3\right)\nonumber\\
&&\times\left[1-i{\rm sgn}(q)\gamma_1\gamma_2\tanh v_2\right]\gamma_\nu\bigg\}
e^{-{(v_1+v_2)(m^2+\omega_n^2+p_z^2)+(\tanh v_1+\tanh v_2)p_\perp^2\over |qB|}}
\end{eqnarray}
with the summation and integration $\sum_{n{\bm p}v_1v_2}=\sum_{n=-\infty}^\infty\int d^3{\bm p}/(2\pi)^3\int_0^\infty dv_1 dv_2$, where the Matsubara summation is over the quark frequency $\omega_n=-ip_0=(2n+1)\pi T$. Using the exchange symmetry between $v_1$ and $v_2$ and computing the trace in Dirac space give
\begin{eqnarray}
\Pi_{\mu\nu}(T,B) &=& {2g^2T\over |qB|^2}\sum_{n{\bm p}v_1v_2}\bigg\{{g_{\mu\nu}p_\perp^2+2(p-p_{||})_\mu(p-p_{||})_\nu\over\cosh^2 v_1\cosh^2 v_2}\nonumber\\
&& +m^2\left[g_{\mu\nu}+(g^{||}_{\mu\nu}-g^\perp_{\mu\nu})\tanh v_1\tanh v_2\right]+\omega_n^2\left[g^\perp_{\mu\nu}-\delta^{||}_{\mu\nu}-(g^\perp_{\mu\nu}+\delta^{||}_{\mu\nu})\tanh v_1\tanh v_2\right]\nonumber\\
&&+p_z^2\left[g^\perp_{\mu\nu}+\delta^{||}_{\mu\nu}-(g^\perp_{\mu\nu}-\delta^{||}_{\mu\nu})\tanh v_1\tanh v_2\right]\bigg\}e^{-{(v_1+v_2)(m^2+\omega_n^2+p_z^2)+(\tanh v_1+\tanh v_2)p_\perp^2\over |qB|}}
\end{eqnarray}
with the definitions of $p_\perp^2 = p_x^2+p_y^2,\ g^{||}_{\mu\nu} = \textrm{diag}(1,0,0,-1),\ g^\perp_{\mu\nu} = \textrm{diag}(0,-1,-1,0),\ \delta^{||}_{\mu\nu} = \textrm{diag}(1,0,0,1)$ and $\delta^\perp_{\mu\nu} = \textrm{diag}(0,1,1,0)$.

It is easy to see that all the off-diagonal elements ($\mu\neq\nu$) of the polarization vanish automatically, and we need to consider the diagonal elements only. We can further divide the diagonal polarization into a parallel and a perpendicular parts $\Pi_{\mu\mu}^{||}$ with $\mu\in\{0,3\}$ and $\Pi_{\mu\mu}^\perp$ with $\mu\in\{1,2\}$. Let's first calculate the parallel part which is directly related to the color screening mass, see the next section. Taking into account the rotational symmetry around the $z-$axis, the parallel polarization becomes
\begin{eqnarray}
\Pi_{\mu\mu}^{||}(T,B) &=& {2g^2T\over |qB|^2}\sum_{n{\bm p}v_1v_2}\left\{{g^{||}_{\mu\mu}p_\perp^2\over\cosh^2v_1\cosh^2v_2}
+\left(1+\tanh v_1\tanh v_2\right)\left[\delta^{||}_{\mu\mu}\left(-\omega_n^2+p_z^2\right)+g^{||}_{\mu\mu}m^2\right]\right\}\nonumber\\
&&\times e^{-{(v_1+v_2)(m^2+\omega_n^2+p_z^2)+(\tanh v_1+\tanh v_2)p_\perp^2\over |qB|}}.
\end{eqnarray}

Introducing functions $d_n(\alpha)$ defined through the Legendre functions $d_n(\alpha)=(-1)^n e^{-\alpha}\left[L_n(2\alpha)-L_{n-1}(2\alpha)\right]$ with $L_{-1}(2\alpha)=0$, the completeness relation $\sum_{n=0}^\infty d_n(\alpha)e^{-2inv}=e^{-i\alpha\tanh v}$ can alternatively be expressed as~\cite{Chyi} 
\begin{eqnarray}
\label{complete}
&& \sum_{n=0}^\infty d_n(\alpha)e^{-2nv} = e^{-\alpha\tanh v},\nonumber\\
&& \sum_{n=0}^\infty 2n d_n(\alpha)e^{-2nv} = {\alpha\over\cosh^2 v}e^{-\alpha\tanh v},\nonumber\\
&& \sum_{n=0}^\infty d_n'(\alpha)e^{-2nv} = -\tanh v e^{-\alpha\tanh v}
\end{eqnarray}
by taking the replacement of $v$ by $-iv$. Choosing $\alpha = p_\perp^2/|qB|$ and expressing $\cosh v_i$ and $\tanh v_i$ with $i=1,2$ by the above summations, the parallel polarization can be written as
\begin{eqnarray}
\Pi_{\mu\mu}^{||}(T,B) &=& {2g^2T\over |qB|^2}\sum_{n{\bm p}v_1v_2n_1n_2}\bigg\{4|qB|g^{||}_{\mu\mu}{n_1 n_2d_{n_1}(\alpha)d_{n_2}(\alpha)\over \alpha}+\left[d_{n_1}(\alpha)d_{n_2}(\alpha)+d_{n_1}'(\alpha)d_{n_2}'(\alpha)\right]\nonumber\\
&&\times\left[\delta^{||}_{\mu\mu}\left(-\omega_n^2+p_z^2\right)+g^{||}_{\mu\mu}m^2\right]\bigg\} e^{-2(n_1v_1+n_2v_2)-(v_1+v_2){m^2+\omega_n^2+p_z^2\over |qB|}}
\end{eqnarray}
with the summation $\sum_{n_1,n_2=0}^\infty$.

We then change the transverse momentum integration to $\alpha-$integration, the rotational symmetry in the transverse plane leads to $\int d^3{\bm p}/(2\pi)^3 = \int_{-\infty}^\infty dp_z/(2\pi)^2\int_0^\infty dp_\perp p_\perp = \int_{-\infty}^\infty dp_z/(2\pi)^2|qB|/2\int_0^\infty d\alpha$. Using the orthogonal relations for the functions $d_n$,
\begin{eqnarray}
\label{orthogonal}
&& \int_0^\infty d\alpha {n_1n_2\over\alpha}d_{n_1}(\alpha)d_{n_2}(\alpha)=n_1\delta_{n_1n_2},\nonumber\\
&& \int_0^\infty d\alpha \left[d_{n_1}(\alpha)d_{n_2}(\alpha)+d_{n_1}'(\alpha)d_{n_2}'(\alpha)\right] = \left(2-\delta_{n_10}\right)\delta_{n_1n_2},
\end{eqnarray}
the integration over $v_1$, $v_2$ and $\alpha$ gives
\begin{eqnarray}
\Pi_{\mu\mu}^{||}(T,B) &=& g^2T|qB|\sum_{np_zn_1n_2}{1\over m^2+\omega_n^2+p_z^2+2n_1|qB|}{1\over m^2+\omega_n^2+p_z^2+2n_2|qB|}\nonumber\\
&&\times\bigg\{4g^{||}_{\mu\mu}n_1|qB|+\left(2-\delta_{n_10}\right)\left[\delta^{||}_{\mu\mu}\left(-\omega_n^2+p_z^2\right)+g^{||}_{\mu\mu}m^2\right]
\bigg\}\delta_{n_1n_2}
\end{eqnarray}
with the longitudinal integration $\sum_{p_z}=\int dp_z/(2\pi)^2$. Doing the summation over $n_2$ analytically and employing the derivative relation,
\begin{equation}\label{complete_differential}
{\partial\over\partial p_z}\left({p_z\over m^2+\omega_n^2+p_z^2+2n_1|qB|}\right) = {m^2+\omega_n^2-p_z^2+2n_1|qB|\over\left(m^2+\omega_n^2+p_z^2+2n_1|qB|\right)^2},
\end{equation}
the parallel polarization is finally simplified as
\begin{equation}
\label{parallel}
\Pi_{\mu\mu}^{||}(T,B) = g^2T|qB|\sum_{np_zn_1}{\left(2-\delta_{n_10}\right)(\delta^{||}_{\mu\mu}+g^{||}_{\mu\mu})(-\omega_n^2+p_z^2)\over(m^2+\omega_n^2+p_z^2+2n_1|qB|)^2}.
\end{equation}

The physics of the positive integer $n_1$ becomes now clear. It is well-known that in quantum mechanics the transverse Landau energy levels of a quark propagating in the external magnetic field are
\begin{equation}
\label{Landau}
\epsilon_{n_1}^2 = 2n_1|qB|,\ \ \ \ \ n_1=0,1,\cdots,\infty.
\end{equation}

We now turn to the calculation of the perpendicular polarization
\begin{eqnarray}
\Pi_{ii}^\perp(T,B) &=& {2g^2T\over |qB|^2}\sum_{n{\bm p}v_1v_2}
\left[{-p_\perp^2+2p_i^2\over\cosh^2v_1\cosh^2v_2}-\left(1-\tanh v_1\tanh v_2\right)\left(\omega_n^2+p_z^2+m^2\right)\right]\nonumber\\
&&\times e^{-{(v_1+v_2)(m^2+\omega_n^2+p_z^2)+(\tanh v_1+\tanh v_2)p_\perp^2\over |qB|}}
\end{eqnarray}
with $i=1,2$ and $p_1=p_x, p_2=p_y$. When the magnetic field disappears, it is easy to check that the perpendicular polarization vanishes automatically,
\begin{equation}
\Pi_{ii}^\perp(T,0) = 2g^2T\sum_{n{\bm p}}{-\omega_n^2+2p_i^2-{\bm p}^2-m^2\over (\omega_n^2+{\bm p}^2+m^2)^2} = -2g^2T\sum_{n{\bm p}}{\partial\over\partial p_i}{p_i\over \omega_n^2+{\bm p}^2+m^2} =0.
\end{equation}
This comes back to the known result at finite temperature~\cite{QFT,Laine:2006ns,Schneider:2003uz}.

To see the magnetic field effect we consider the difference between the two cases with and without magnetic field. From the exchange symmetry between $p_x$ and $p_y$, the difference in the perpendicular polarization can be expressed as
\begin{eqnarray}
\delta\Pi_{ii}^\perp (T,B)&=& \Pi_{ii}^\perp(T,B)-\Pi_{ii}^\perp(T,0)\nonumber\\
&=&{2g^2T\over |qB|^2}\sum_{n{\bm p}v_1v_2}
\bigg\{\left(\omega_n^2+p_z^2+m^2\right)\bigg[e^{-(v_1+v_2){p_\perp^2\over |qB|}}\nonumber\\
&&-\left(1-\tanh v_1\tanh v_2\right)e^{-(\tanh v_1+\tanh v_2){p_\perp^2\over |qB|}}\bigg]\bigg\}e^{-(v_1+v_2){m^2+\omega_n^2+p_z^2\over |qB|}}.
\end{eqnarray}
Similar to the treatment for the parallel part, we again introduce the variable $\alpha=p_\perp^2/|qB|$ and the sums over $n_1$ and $n_2$ via using the completeness relations (\ref{complete}). Then by integrating out $v_1, v_2$ and $\alpha$ and using the orthogonal relation
\begin{equation}
\int_0^\infty d\alpha \left[d_{n_1}(\alpha)d_{n_2}(\alpha)-d_{n_1}'(\alpha)d_{n_2}'(\alpha)\right] = \delta_{|n_1-n_2|,1},
\end{equation}
the difference becomes
\begin{eqnarray}
\delta\Pi_{ii}^\perp(T,B) &=& g^2T\sum_{np_zn_1n_2}
\left(\omega_n^2+p_z^2+m^2\right)\nonumber\\
&&\times\bigg[{\delta_{n_10}\delta_{n_20}\over m^2+\omega_n^2+p_z^2}-{|qB|\delta_{|n_1-n_2|,1}\over \left(m^2+\omega_n^2+p_z^2+2n_1|qB|\right)\left(m^2+\omega_n^2+p_z^2+2n_2|qB|\right)}\bigg].
\end{eqnarray}
Taking the relation on the summation over $n_1$ and $n_2$ for any constant $\lambda$,
\begin{eqnarray}
\sum_{n_1,n_2=0}^\infty{\delta_{|n_1-n_2|,1}\over\left(\lambda+2n_1\right)\left(\lambda+2n_2\right)} &=& \sum_{n_2>n_1\geq 0}{\delta_{|n_1-n_2|,1} \over n_2-n_1}\left({1\over\lambda+2n_1}-{1\over\lambda+2n_2}\right)\nonumber\\
&=& \sum_{n_1\geq 0}\left({1\over\lambda+2n_1}-{1\over\lambda+2n_1+2}\right) = {1\over \lambda},
\end{eqnarray}
the difference vanishes,
\begin{eqnarray}
\delta\Pi_{ii}^\perp(T,B) &=& g^2T\sum_{np_z}\left(\omega_n^2+p_z^2+m^2\right)
\left({1\over m^2+\omega_n^2+p_z^2}-{1\over m^2+\omega_n^2+p_z^2}\right)\nonumber\\
&=& 0.
\end{eqnarray}
Therefore, there is no perpendicular polarization in any case with and without magnetic field, $\Pi_{\mu\mu}^\perp(0)=0$.

\section{Color screening mass}
 At one-loop level, the gluon propagator is controlled by not only the quark loop but also the gluon loop and ghost loop. Since gluons and ghosts do not carry charge, they are not coupled to the external magnetic field, and the temperature dependence of the gluon and ghost induced gluon polarization $\overline\Pi_{\mu\nu}(k)$ is well investigated in literatures~\cite{QFT}. After resummation over the quark loops, gluon loops and ghost loops, one derives the total gluon propagator and in turn the total screening mass
\begin{eqnarray}
&& m_D^2(T,B) = m_Q^2(T,B)+m_G^2(T),\nonumber\\
&& m_Q^2(T,B)=-\Pi_{00}^{||}(T,B),\nonumber\\
&& m_G^2(T)=-\overline\Pi_{00}^{||}(T).
\end{eqnarray}

The gluon and ghost loop induced screening mass $m_G^2(T)$ which is independent of electromagnetic interaction can be taken from Ref.~\cite{Huang:2021ysc},
\begin{equation}
m_G^2(T)={N_c\over 3}g^2T^2,
\end{equation}
and the quark loop induced screening mass $m_Q^2(T,B)$ is controlled by the parallel polarization $\Pi_{00}^{||}$, 
\begin{equation}
m_Q^2(T,B) = -g^2T|qB|\sum_{np_zn_1}\left[\left(2-\delta_{n_1,0}\right){m^2-\omega_n^2+p_z^2+2n_1|qB|\over(m^2+\omega_n^2+p_z^2+2n_1|qB|)^2}\right].
\end{equation}
Considering the Landau energy levels as the quark transverse momentum $p_\perp^2=2n_1|qB|$, and using the trace computation,
\begin{equation}
\textrm{Tr}\left(\gamma_0{1\over\gamma\cdot p-m}\gamma_0{1\over\gamma\cdot p-m}\right) = 4{m^2-\omega_n^2+{\bm p}^2\over(m^2+\omega_n^2+{\bm p}^2)^2},
\end{equation}
the summation over the Landau energy levels $\sum_{n_1}$ can be effectively expressed, together with the $p_z$-integration, as a three dimensional integration, 
\begin{equation}
m_Q^2(T,B) = -{g^2\over 2}T\sum_{p_0{\bm p}}\textrm{Tr}\left(\gamma_0{1\over\gamma\cdot p-m}\gamma_0{1\over\gamma\cdot p-m}\right)\rho_B(p_\perp^2),
\end{equation}
where $\rho_B$ is the magnetic field controlled transverse momentum distribution,
\begin{equation}
\rho_B(p_\perp^2) = |qB|\sum_{n_1=0}^\infty\left(2-\delta_{n_1,0}\right)\delta(p_\perp^2-2n_1|qB|),
\end{equation}
the $\delta$-function means the Landau quantization: quarks are confined on the quantum orbit in phase space $p_\perp^2=2n_1qB$.

It can be proven that this general screening mass covers the known result in the limit of weak magnetic field. When the magnetic field disappears, the summation over the Landau levels becomes an integration, according to the Riemann summation rule,
\begin{equation}
\lim_{B\rightarrow 0}\rho_B(p_\perp^2) = \int_0^\infty d\xi\ \delta(p_\perp^2-\xi)=1,
\end{equation}
we therefore go back to the familiar screening mass as a function of temperature for massless quarks~\cite{Huang:2021ysc}
\begin{equation}
m_Q^2(T,0)={N_f\over 6}g^2T^2,
\end{equation}
where $N_f$ comes from the flavor summation.

We now subtract the pure temperature effect from the screening mass to focus on the magnetic field induced mass shift,
\begin{eqnarray}\label{pure_magnetic_contribution}
\delta m_D^2(T,B) &=& m_D^2(T,B)-m_D^2(T,0)\nonumber\\
&=& -{g^2\over 2}T\sum_{p_0{\bm p}}\textrm{Tr}\left(\gamma_0{1\over\gamma\cdot p-m}\gamma_0{1\over\gamma\cdot p-m}\right)\left[\rho_B(p_\perp^2)-1\right].
\end{eqnarray}

For massless quarks, by summing up the Landau levels, one obtains the Taylor expansion of the screening mass in terms of $|qB|$ in the limit of weak magnetic field,
\begin{eqnarray}\label{weak_expansion}
\delta m_D^2(T,B) &=& -g^2T\sum_f\left[\sum_{np_z}\left({p_z^2-\omega_n^2\over(p_z^2+\omega_n^2)^2}|q_fB|
+{4\over 3}{p_z^2-\omega_n^2\over (p_z^2+\omega_n^2)^3}|q_fB|^2\right)+\mathcal{O}(|q_fB|^4)\right]\nonumber\\
&=& \sum_f\left[{7\zeta(3)\over 48\pi^4}{g^2\over T^2}|q_fB|^2+\mathcal{O}(|q_fB|^4)\right].
\end{eqnarray}
Here we have considered the contribution from different flavors and the flavor dependence of the quark charge $q\to q_f$ in the quark loop calculation. This result agrees with the one derived in Refs.\cite{Hasan:2020iwa,Karmakar:2018aig}. It is straightforward to calculate the corrections from higher orders.

For the other limit of strong magnetic field, we can take only the lowest Landau level ($n_L=0$). For massless quarks we analytically obtain
\begin{equation}\label{strong_magnetic_field}
m_Q^2(T,B) = {g^2\over 4}\sum_f{|q_fB|\over T}\sum_{p_z}{1\over \cosh^2\left(|p_z|/(2T)\right)} = {g^2\over 4\pi^2}\sum_f|q_fB|.
\end{equation}
This result is exactly what people derived previously~\cite{Bonati:2017uvz,Singh:2017nfa}. It is straightforward to consider the correction from higher Landau levels to the screening mass in our frame.

We now generally calculate the mass shift $\delta m_D^2(T,B)$ without considering any restriction to the temperature and magnetic field. Again we consider massless quarks. Summing up all the Landau levels in the quark loop induced polarization (\ref{parallel}) leads to
\begin{equation}
\delta m_D^2(T,B) = g^2T\sum_{np_zf}{p_z^2-\omega_n^2\over|q_fB|}{\mathcal K}\left({p_z^2+\omega_n^2\over 2|q_fB|}\right),
\end{equation}
where the function ${\mathcal K}$ is defined as ${\mathcal K}(x) =x^{-2}/2+x^{-1}-\psi'(x)$ with $\psi(x)=\Gamma'(x)/\Gamma(x)$. Note that, in the Taylor expansion of the mass shift in terms of $|q_fB|$, the linear term disappears automatically, see (\ref{weak_expansion}), we can safely subtract $x^{-2}/2$ from ${\mathcal K}(x)$. Taking the integrated function as ${\mathcal K}(x)-x^{-2}/2$ and doing partial integration, we have
\begin{eqnarray}
{p_z^2-\omega_n^2\over|q_fB|}\left[{\mathcal K}\left({p_z^2+\omega_n^2\over 2|q_fB|}\right)-{2|q_fB|^2\over \left(p_z^2+\omega_n^2\right)^2}\right]
&=& 2\int_0^\infty d\xi\left[1-{|q_fB|\xi^2/(2\pi^2T^2)\over 1-e^{-{|q_fB|\xi^2\over 2\pi^2T^2}}}\right]e^{-{2p_z^2\xi^2\over 4\pi^2T^2}}{\partial\over\partial\xi}e^{{p_z^2-\omega_n^2\over 4\pi^2 T^2}\xi^2}\nonumber\\
&=& 2\int_0^\infty d\xi e^{-{p_z^2+\omega_n^2\over 4\pi^2T^2}\xi^2}\left(-{\partial\over\partial\xi}+{p_z^2\over\pi^2T^2}\xi\right)\left[1-{|q_fB|\xi^2/(2\pi^2T^2)\over 1-e^{-{|q_fB|\xi^2\over 2\pi^2 T^2}}}\right],
\end{eqnarray}
which results in  
\begin{equation}
{p_z^2-\omega_n^2\over |q_fB|}{\mathcal K}\left({p_z^2+\omega_n^2\over 2|q_fB|}\right) = 2\int_0^\infty d\xi e^{-{p_z^2+\omega_n^2\over 4\pi^2T^2}\xi^2}\left(-{\partial\over\partial\xi}+{p_z^2\over\pi^2 T^2}\xi\right)\left[1-{|q_fB|\xi^2\over 4\pi^2T^2}\coth\left({|q_fB|\xi^2\over 4\pi^2T^2}\right)\right].
\end{equation}
Now we can analytically sum up the Matsubara frequency and integrate the longitudinal momentum, the mass shift is finally written as
\begin{equation}
\delta m_D^2(T,B) ={2g^2T^2\over\pi^{1/2}}\sum_f\int_0^\infty d\xi {\vartheta_2(0,e^{-\xi^2})\over \xi^2}{\mathcal M}\left({|q_fB|\xi^2\over 4\pi^2T^2}\right),
\end{equation}
where $\vartheta _2$ is the elliptic theta function $\vartheta _2(u,x) = 2x^{1/4}\sum_{n=0}^\infty x^{n(n+1)}\cos[(2n+1)u]$, and ${\mathcal M}$ is defined as ${\mathcal M}(x) = 1-x^2/\sinh^2 x$. Considering the relations $\vartheta_2(0,e^{-\xi^{2}})=\sqrt{\pi}/\xi\vartheta_{4}(0,e^{-\pi^{2}/\xi^{2}})$ and $\vartheta_{2}(0,e^{-\xi^{2}})\approx\sqrt{\pi}/\xi$ in the limit $\xi\rightarrow0^{+}$ which corresponds to the limit of strong magnetic field, we obtain  
\begin{eqnarray}
	\delta m_D^2(T,B) ={g^{2}\over4\pi^{2}}\sum_f|q_{f}B|+{2g^2T^2\over\pi^{1/2}}\sum_f\int_0^\infty d\xi {\vartheta_2(0,e^{-\xi^2})-\sqrt{\pi}/\xi\over \xi^2}{\mathcal M}\left({|q_fB|\xi^2\over 4\pi^2T^2}\right).
\end{eqnarray}

\begin{figure}[htbp]
	\includegraphics[width=10.0cm]{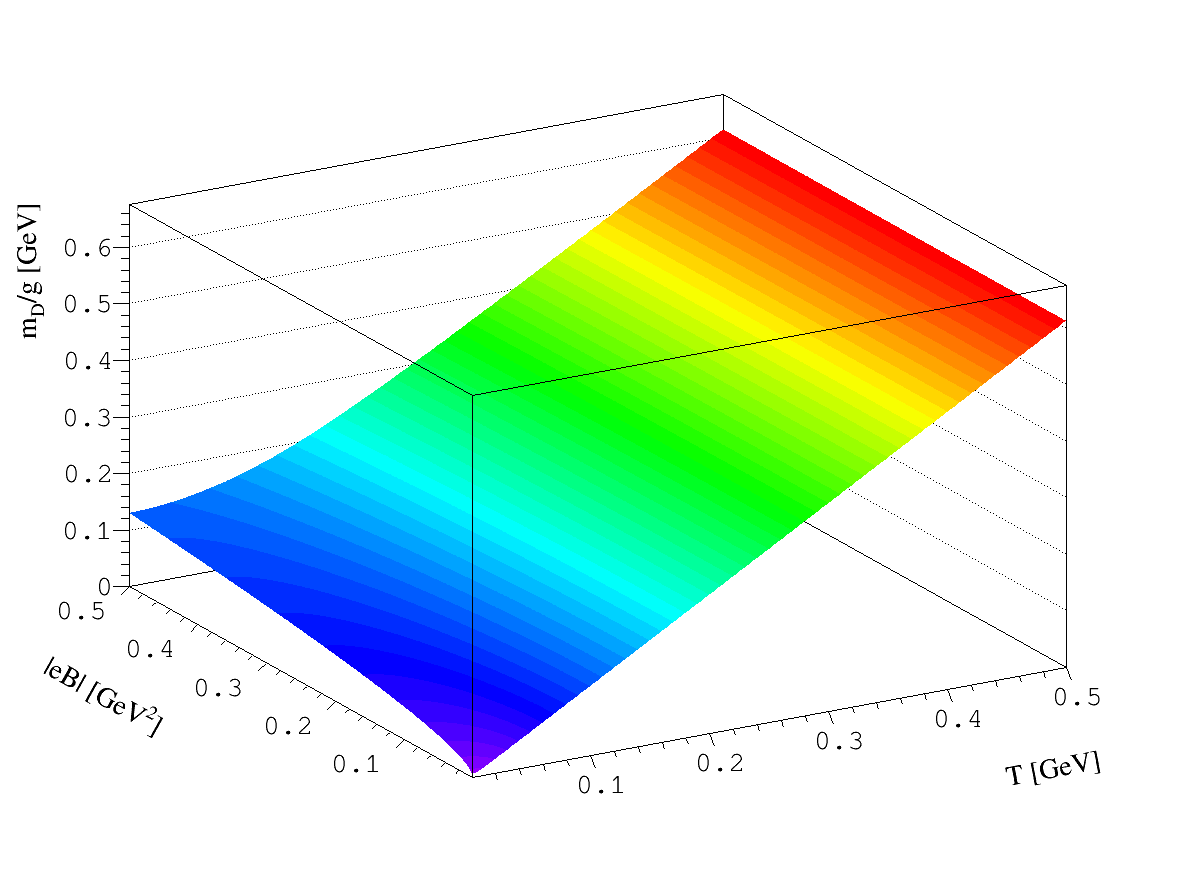}
	\caption{The total Debye screening mass scaled by the coupling constant $m_D(T,B)/g$ as a function of temperature $T$ and external magnetic field strength $|eB|$. }
	\label{fig1}
\end{figure}

\begin{figure}[htbp]
\includegraphics[width=10.0cm]{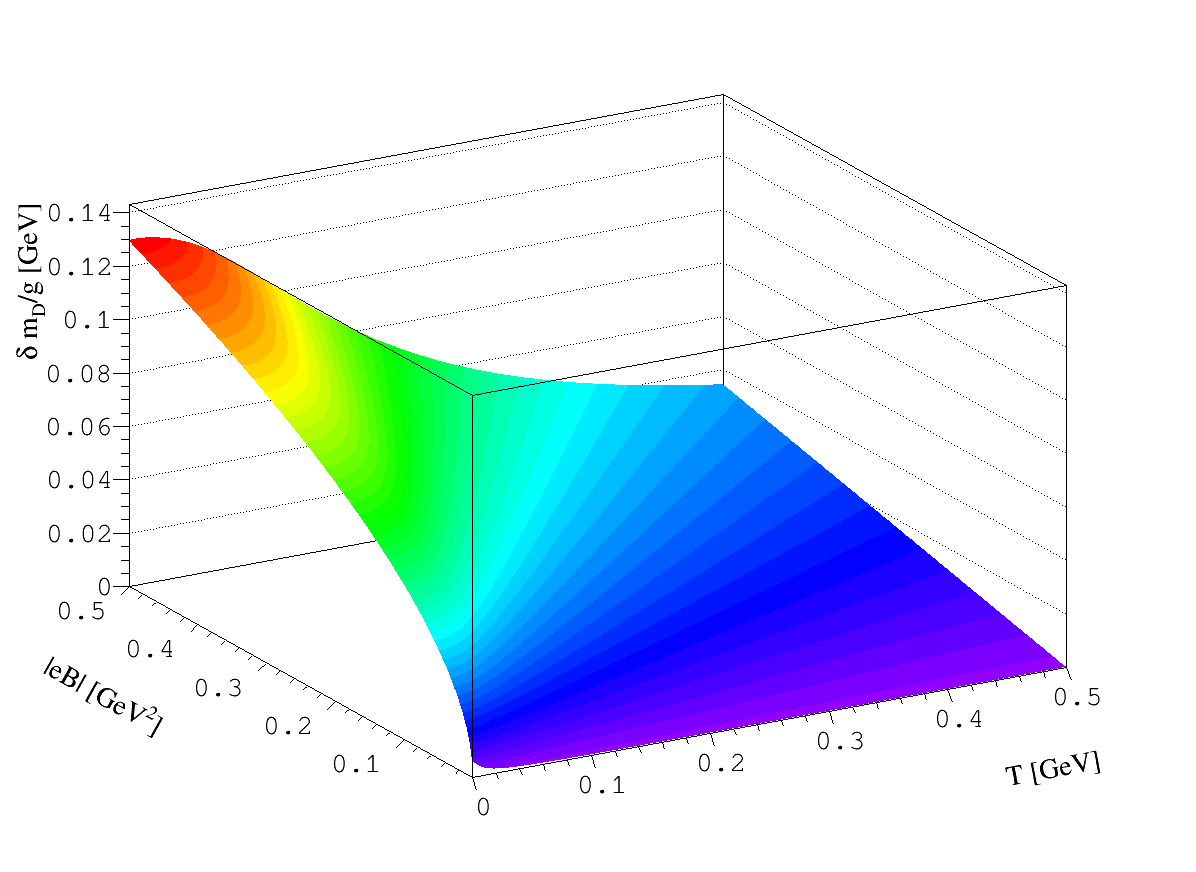}
\caption{The magnetic field induced mass shift scaled by the coupling constant $\delta m_D(T,B)/g$ as a function of temperature $T$ and external magnetic field strength $|eB|$. }
\label{fig2}
\end{figure}

The numerical results for the Debye screening mass $m_D(T,B)$ and the mass shift $\delta m_D(T,B)=\sqrt{m_D^2(T,B)-m_D^2(T,0)}$ are shown in Figs.\ref{fig1} and \ref{fig2} as functions of $T$ and $|eB|$. In the frame of one-loop resummation, the screening mass square is at the order of $g^2$, and therefore the scaled mass $m_D/g$ and mass shift $\delta m_D/g$ are coupling constant independent. In the hot and magnetized medium created in high energy nuclear collisions at LHC, the screening mass induced by temperature $m_D(T,0)/g=\sqrt{3/2}T\sim 0.6$ GeV at $T=0.5$ GeV is much larger than the one by magnetic field $m_D(0,B)/g = 0.13$ GeV at $eB=0.5$ GeV $^2\sim 25 m_\pi^2$. With increasing temperature, the broken translation invariance caused by the magnetic field is gradually restored by the thermal motion, and the mass shift drops down continuously.   

\begin{figure}[htbp]
	\includegraphics[width=10.0cm]{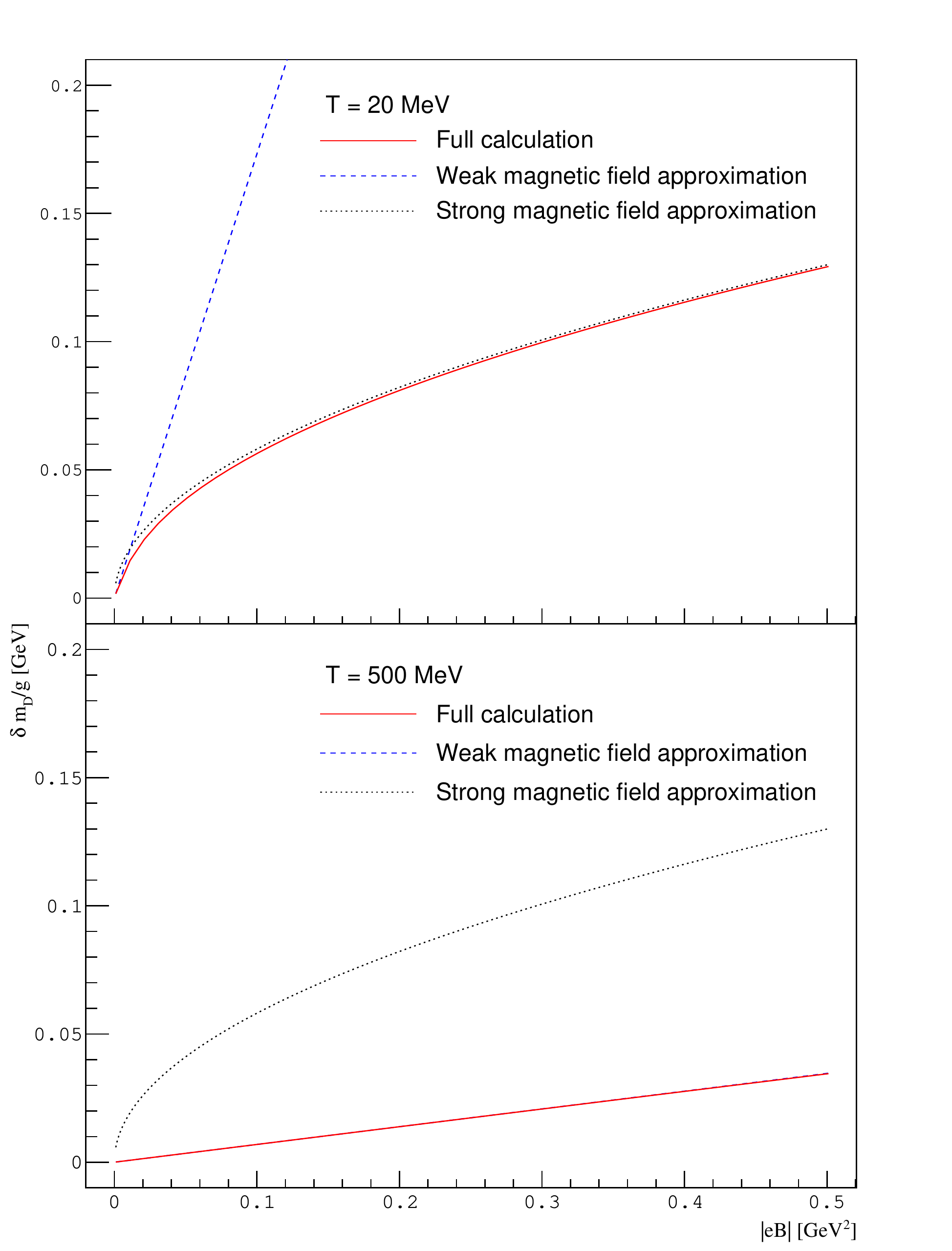}
	\caption{The comparison of the full calculation (solid lines) with the limits of weak (dashed lines) and strong (dotted lines) magnetic field at low temperature $T=20$ MeV (upper panel) and high temperature $T=500$ GeV (lower panel). }
	\label{fig3}
\end{figure}

We now take numerical comparison of our full calculation with the approximations of weak (Eq.(\ref{weak_expansion})) and strong (Eq.(\ref{strong_magnetic_field})) magnetic field in Fig.\ref{fig3}, where the weak and strong limits are relative to the medium temperature. In order to make our comparison meaningful, we take two temperatures, $T=20$ MeV corresponding to the cold quark matter in the core of compact stars and $T=500$ MeV corresponding to the initial fireball in high energy nuclear collisions at RHIC and LHC energies where the created magnetic field is the strongest. It is clear that for compact stars the magnetic field effect is essential and the limit of strong magnetic field including only the lowest Landau level is a good approximation. While the magnetic field created in the initial stage of heavy ion collisions is extremely strong, its effect on the hot QCD matter can safely be considered as a perturbation with respect to the initial fireball temperature.        

\section{Summary}
The color interaction is screened in QCD matter at finite temperature and further suppressed in external magnetic field. We calculated in this paper the color screening mass in the frame of resummed perturbative QCD theory without restriction to the magnetic field strength. In the quark loop calculation, the Landau energy levels $\epsilon_n^2=2n|qB|$ for the propagating quark and anti-quark are naturally embedded into the screening mass. Our full calculation covers the often used limit of weak magnetic field at high temperature and the limit of strong magnetic field at low temperature. While the magnetic field created in high energy nuclear collisions at RHIC and LHC energies is perhaps the strongest in nature, its effect on the formed quark-gluon plasma is still weaker in comparison with the temperature effect and can safely be considered as a perturbation. 

{\bf Acknowledgement}: The work is supported by the Guangdong Major Project of Basic and Applied Basic Research No. 2020B0301030008 and the NSFC grants Nos. 11890712 and 12075129.

\end{document}